\documentclass{PoS}

\title{Dynamical simulations of QCD at finite temperature with a truncated perfect action\footnote{Preprint BI-TP 2006/35}}

\ShortTitle{Dynamical simulations of QCD at finite temperature
with a truncated perfect action}

\author{\speaker{Stanislav Shcheredin} and Edwin Laermann \\
        Fakult\"at f\"ur Physik, Universit\"at Bielefeld,\\ D-33615 Bielefeld, Germany \\
        E-mail: \email{shchered@physik.uni-bielefeld.de},
        \email{edwin@physik.uni-bielefeld.de}}

\abstract{The Hypercube operator determines a variant of the
approximate, truncated perfect fermion action. In this pilot study
we are going to report on first experiences in dynamical QCD
simulations with the Hypercube fermions. We apply this formulation
in an investigation of the finite temperature transition for two
flavours. On lattices of size $8^3\times 4$ we explore the phase
diagram. Physical scales are estimated from pseudoscalar and
vector meson masses obtained on $8^3\times 16$ lattices. We observe the presence of a metastability region
but do not find evidence for an Aoki phase. The
Hypercube operator allows us to simulate at ratios of pseudoscalar
to vector meson masses at least as small as $0.8$ at the thermal
crossover at $N_t=4$, which renders this formulation cheaper than
the Wilson like fermions.}

\FullConference{XXIVth International Symposium on Lattice Field Theory\\
                July 23-28, 2006\\
                Tucson, Arizona, USA}

\begin{document}

\section{Motivation}
There are infinitely many ways to discretize the fermionic action
of QCD which give the same naive continuum limit. In the class of
actions preserving flavour symmetry the most attractive choice
from the implementation point of view is the Wilson operator. The
application of the Wilson matrix on a spinor field is very cheap
since only the nearest neighbours contribute to the fermion
action. However when it comes to the physics results this choice
may be not the optimal one. Dynamical simulations with two
flavours of Wilson fermions revealed the Aoki
phase~\cite{Aoki:1983qi} and a first order bulk
transition~\cite{Aoki:2001xq,Farchioni:2004us} on coarse lattices
which are lattice artifacts disappearing in the continuum limit.
Therefore sensible simulations have to be restricted to fine
lattices. This automatically increases the cost of the operator
and makes dynamical simulations at light pion mass with current
computer resources and conventional algorithms prohibitively
expensive. In particular finite temperature simulations at $N_t=4$
revealed that the crossover to the high temperature phase occurs at parameters corresponding to heavy pion masses.

Another approach is the perfect action.  The idea of the perfect
action is based on the iterative application of the Wilson
renormalisation group transformation to some given lattice
formulation starting from a theory defined on a very fine lattice.
After an infinite number of iterations one integrates out all
higher modes of the theory beyond some cutoff and arrives at a
fixed point which constitutes the perfect action. Quantities
calculated with the perfect action are automatically free of any
lattice artifacts i.e. they are in the continuum. Hasenfratz and
Niedermayer  developed the concept of the classically perfect
action~\cite{Hasenfratz:1993sp} where one approximates the Wilson
renormalisation group transformation by a minimisation procedure.
The resulting discretization also has to be truncated to render
the formulation implementable. Still, it contains many more
terms than the conventional ones like Wilson fermions and Wilson
gauge action. Therefore the matrix multiplication will be much
more expensive. However one may gain back by using it at a coarse
lattice spacing which could render the formulation much cheaper in
the end. In addition the classically perfect action obeys the
Ginsparg-Wilson relation and thus it has an exact chiral symmetry
which is absent for the conventional
formulations~\cite{Hasenfratz:1998ri}.

It is complicated to realise -- especially dynamically -- the
whole procedure of a truncated perfect action. Attempts to
dynamically simulate the classically perfect action are reported
in~\cite{Hasenfratz:2005tt}. In this work we simulate a simplified
version of a truncated perfect action, the Hypercube operator
which was introduced by Bietenholz and
Wiese~\cite{Bietenholz:1995cy}. To construct it one starts from
free fermions where the perfect action can be found analytically.
It still contains a vector term plus a scalar term
\begin{equation}
D_{\rm HF}(x,y)=\gamma_\mu  \rho_\mu(x-y)+ \lambda(x-y)\, .
\end{equation}
 Then in~\cite{Bietenholz:1996pf} a truncation to a unit
Hypercube was introduced which would hopefully still not distort
much the perfect properties of the theory. It has been shown that
even a truncated version leads to an excellent agreement with the continuum
Stefan-Boltzmann law already at $N_t=4,5$ which is by far better
than with the Wilson fermions~\cite{Bietenholz:1996pf}. Next one
proceeds to the interacting case with a simple ansatz of
hyperlinks~\cite{Bietenholz:1996pf,Bietenholz1998wx,Orginos:1997fh}
and one uses the Wilson gauge action. This construction leads to a
matrix multiplication which is around $15$ times more expensive
than with the conventional Wilson operator. However one can win
back a factor of around $5$ in the matrix inversion due to a
smaller maximal eigenvalue, which is around $2$, and hence a
smaller condition number. Therefore one is left with a
computational overhead of a factor around $3-4$ depending whether
the matrix is inverted or the smallest eigenvalues of it are
calculated.

The Hypercube operator is not chirally symmetric and therefore a
fine tuning is needed to approach the chiral limit. To do this one
rescales the link variables like
\begin{equation}
U_{x,\mu} \rightarrow uU_{x,\mu}\, .
\end{equation}
Increasing $u$ towards its critical value corresponds to
decreasing the current quark mass i.e.\ the tuning towards zero
pion mass.

\section{Simulation setup and results}
In this study we simulated dynamically two flavour QCD with a
truncated fermionic perfect action and the standard plaquette
gauge action. The quarks obey periodic boundary conditions in
spatial directions and antiperiodic ones in time. We used the
Hybrid Monte Carlo algorithm with the exact fermionic force. As an
integrator we chose the Sexton-Weingarten integration scheme with
partially suppressed $\delta\tau^3$ errors~\cite{Sexton:1992nu}.

\begin{figure}[h]
\begin{center}
\includegraphics[angle=-90,scale=0.6]{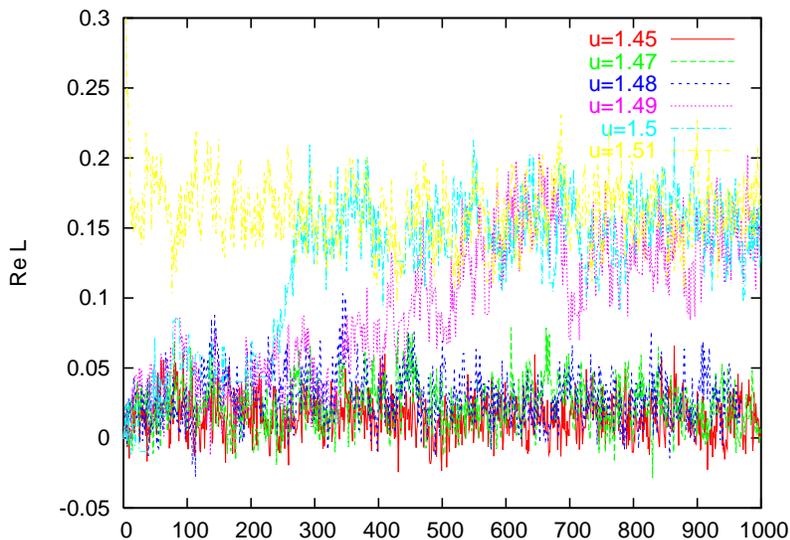}
\end{center}
\caption{Histories of the Polyakov loop at $\beta=5.0$. As the
value of $u$ is increased the system passes from the confined into
the deconfined phase. The transition occurs somewhere between
$u=1.48$ and $1.49$.} \label{figPT}
\end{figure}

We used the Polyakov loop and its susceptibility to monitor the
thermal phase transition in simulations on an $8^3\times4$
lattice. As an example in Figure~\ref{figPT} we show histories of
the Polyakov loop measured over $1000$ trajectories at
$\beta=5.0$. We see that as we increase the value of $u$ (i.e.
decrease the current quark mass) we pass from the confined phase
with zero expectation value of the Polyakov loop to the deconfined
phase indicated by a non-zero Polyakov loop.

\begin{figure}[h]
\begin{center}
\includegraphics[angle=-90,scale=0.6]{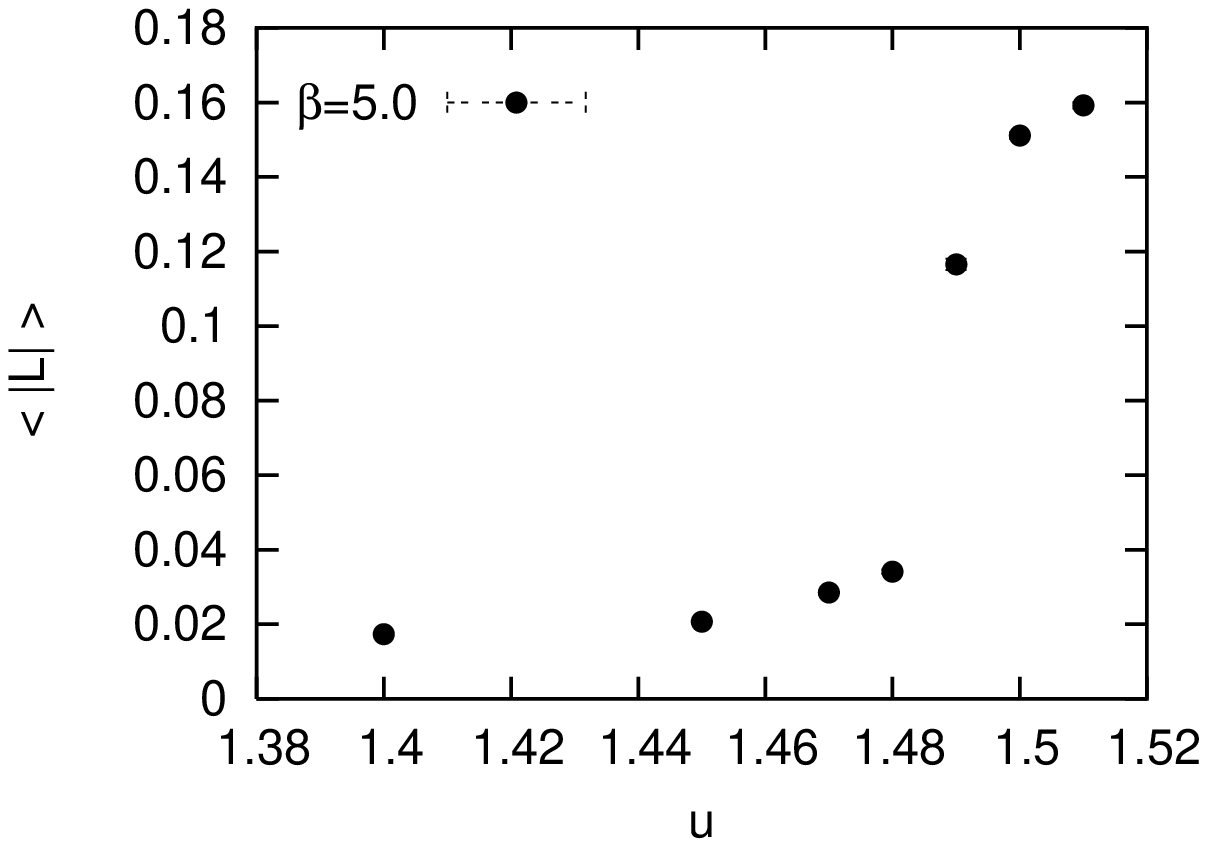}%
\includegraphics[angle=-90,scale=0.6]{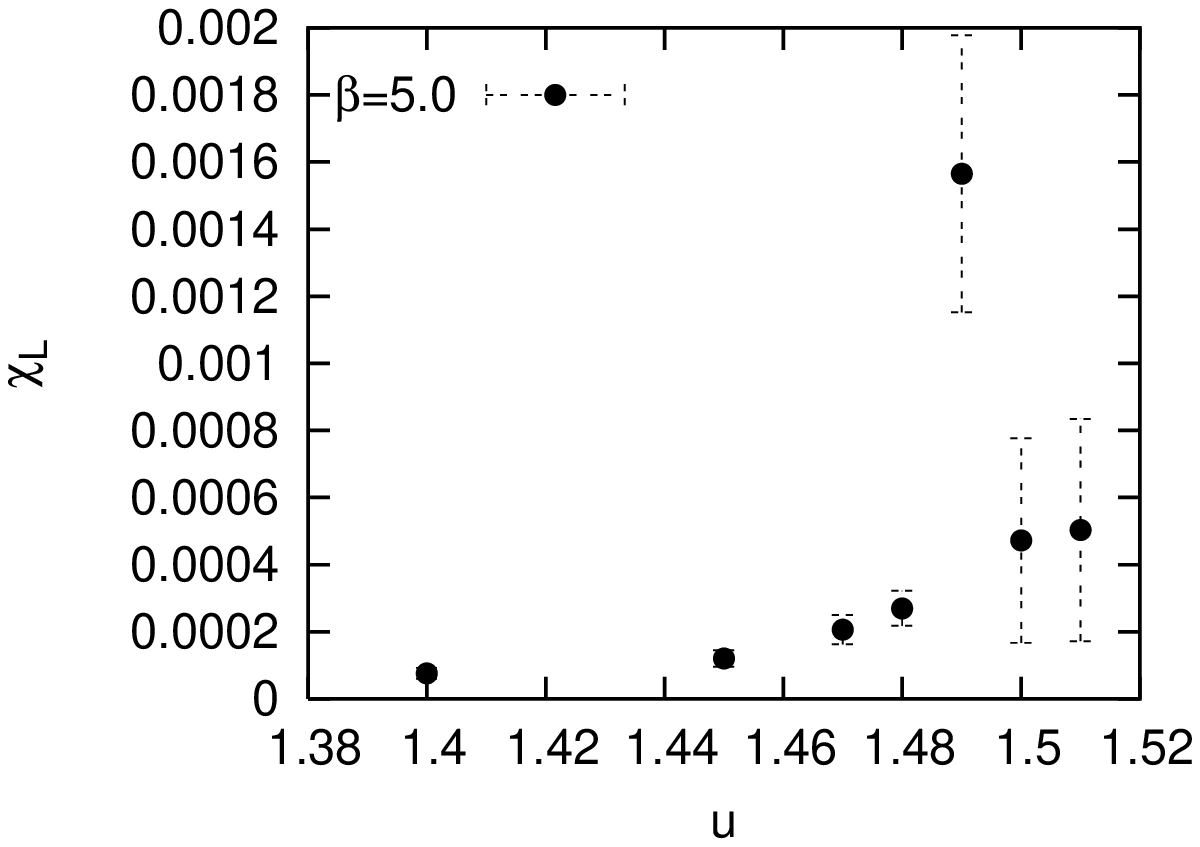}
\end{center}
\caption{The mean value of the Polyakov loop (left) over about
$1000$ trajectories of length $1$ and the susceptibility of the
Polyakov loop (right).} \label{figPVS}
\end{figure}

In Figure~\ref{figPVS} we show an example of the dependence on $u$
of the mean value of the Polyakov loop and its susceptibility at
$\beta=5.0$. We see that the thermal crossover occurs somewhere
between $u=1.48$ and $u=1.49$ as also seen from the histories at
$\beta=5.0$.

\begin{figure}[h]
\begin{center}
\includegraphics[angle=-90,scale=0.6]{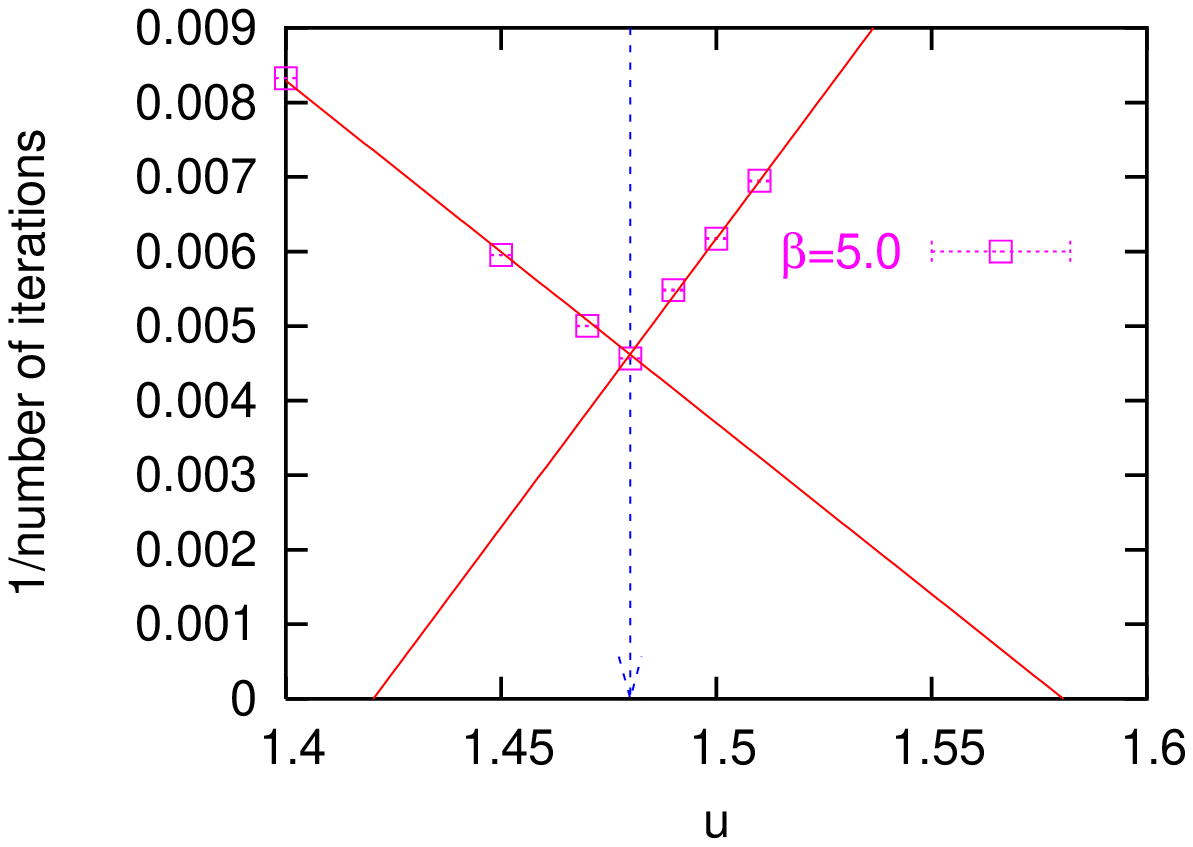}%
\includegraphics[angle=-90,scale=0.6]{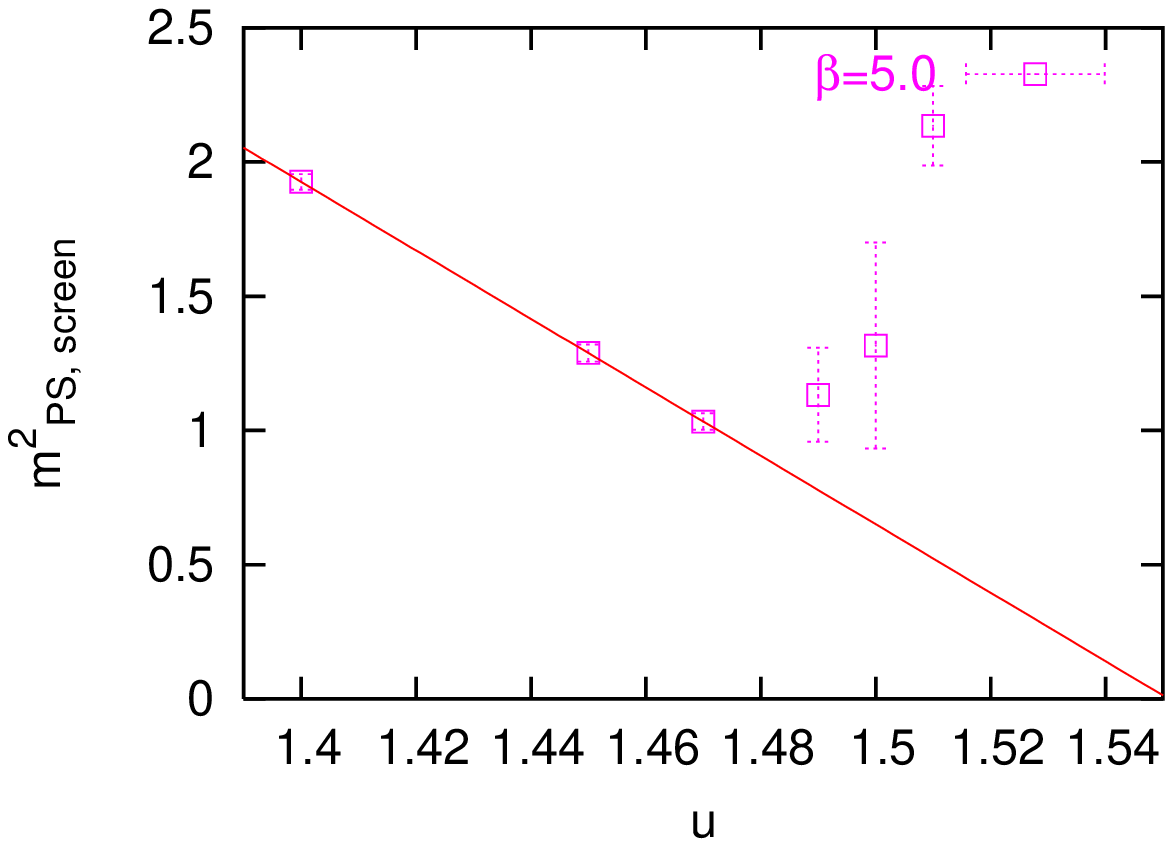}
\end{center}
\caption{On the left plot we show the dependence of the inverse
number of the iterations needed to invert the Hypercube operator
in the course of the molecular dynamics. In particular we see a
nice linear dependence on $u$ in the confined and in the
deconfined regions. This allows us to estimate both the thermal
transition point and the critical point. On the right plot we show
the dependence of the screening pion mass on $u$. This also allows
us to estimate the critical point by an extrapolation from the
confined region to zero.} \label{figNI}
\end{figure}

 We also considered the number of iterations required to invert the Hypercube operator in the course
of the molecular dynamics. This number indicates how singular the
Hypercube operator is i.e. how small the current quark mass is. In
particular in the confined phase we found a nice linear dependence
of the inverse number of iterations on the parameter $u$, see
Figure~\ref{figNI}, up to a $u$ value where the inverse number of
iterations increases linearly again. This point is nearly
identical with the $u$ value where the Polyakov loop
susceptibility develops a peak.
On the same lattice we estimated the critical line by
extrapolating the inverse number of iterations to zero and for
some parameters we also calculated the critical $u$ by
extrapolating the pseudoscalar screening mass to zero, see
Figure~\ref{figNI}. Obviously these two methods suffer from
different systematic errors so that we know only approximately the
critical line.

We also performed zero temperature simulations on $8^3\times 16$
lattices to estimate ratios of the pseudoscalar and the vector
meson masses. Using an empirical formula
\begin{equation}
m_{\rm V}= 756\, {\rm MeV}+450\,{\rm MeV}\left (\frac{m_{\rm
PS}}{m_{\rm V}}\right )^2\, . \label{Eq.S}
\end{equation}
also an estimate of the lattice spacing can be derived from this
ratio.
 For example at $\beta=5.0$ and $u=1.48$ the ratio is around
0.80(1) and the corresponding inverse lattice spacing is $934\pm
11$ MeV. This yields a critical temperature of $T_{\rm C}=234 \pm
3$ MeV. The errors are purely statistical, the systematic error
due to Eq.~(\ref{Eq.S}) can be much bigger.

Next we looked at eigenvalues of the Hypercube operator at
parameters $(\beta,u)$ near to the thermal crossover. We see that
the branches of the eigenvalues are very broad which signals poor
chiral properties at coarse lattices, with lattice spacing around
$0.25-0.35$ fm. This however improves as one proceeds closer to
the critical line where the lattice spacing is decreased.

\begin{figure}[h]
\centering
\includegraphics[angle=-90,scale=0.55]{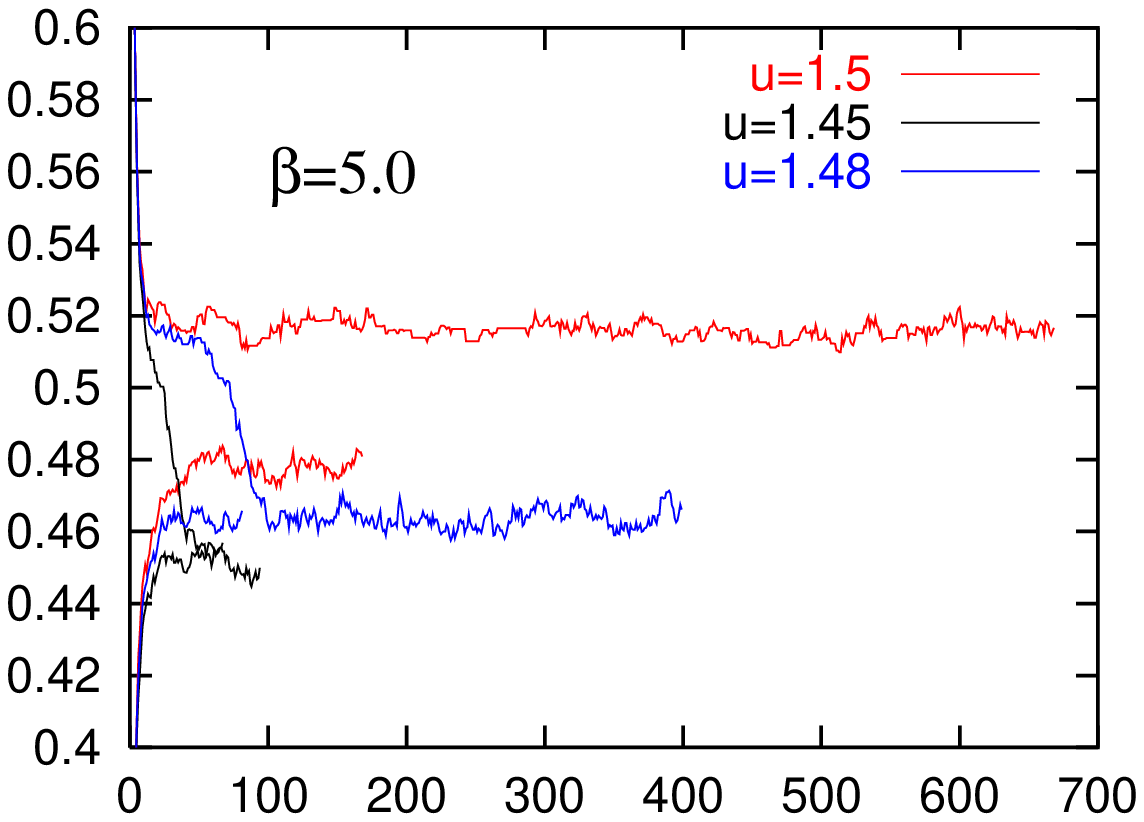}%
\includegraphics[angle=-90,scale=0.55]{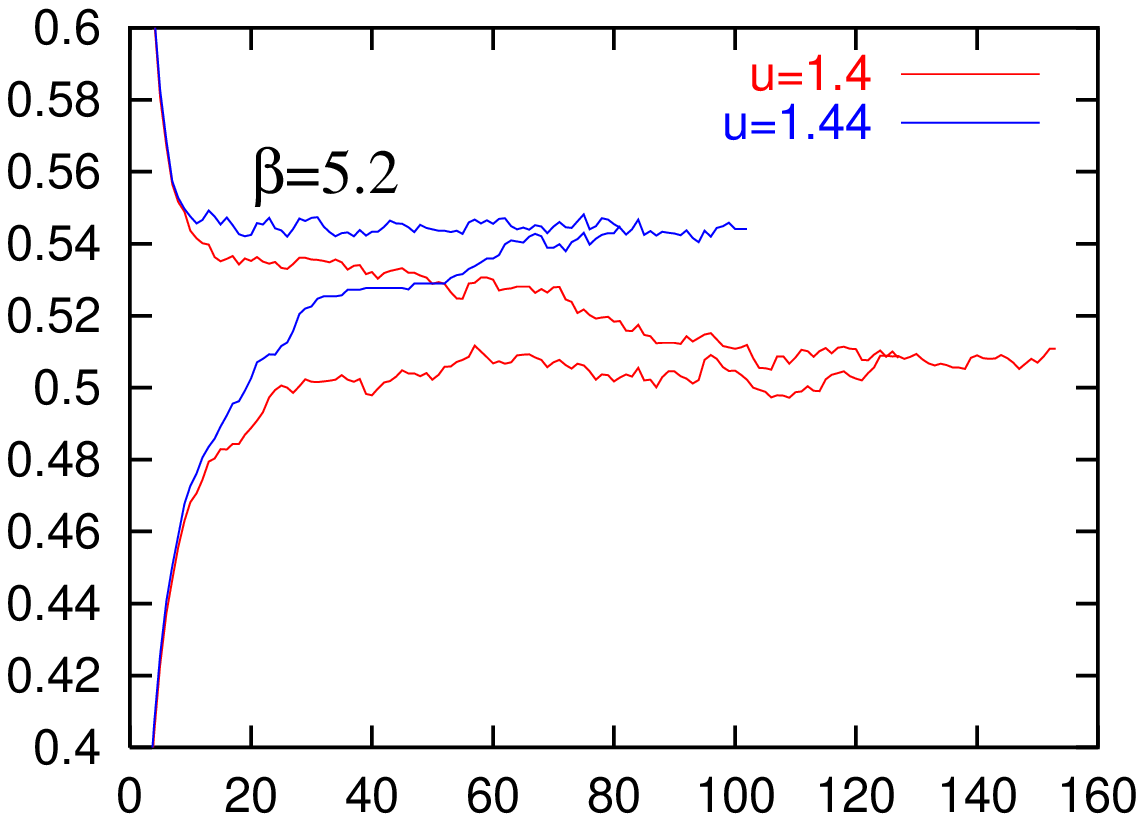}
\caption{On the left plot we show histories of the plaquette at
$\beta=5$, $u=1.45$ and $u=1.5$ in simulations on an $8^3\times
16$ lattice. We see a two-state signal of meta stability at
$u=1.5$ which is absent at $u=1.48$. At $\beta=5.2$ we so far do
not observe such meta stability as shown on the right.}
\label{figPa}
\end{figure}

So far we do not observe the Aoki phase which otherwise would be
signalled by the vanishing pseudoscalar mass in some domain of the
parameter space and result in a singular Hypercube operator and
eventual breakdown of the molecular dynamics.  There is however
another potential problem, namely, the occurrence of a bulk phase
transition.  To monitor this transition we compare runs from hot
and cold starts of the Markov chain. In the vicinity of a first
order phase transition e.g.\ plaquettes  level out at different
values. Our present simulations indicate such a two state signal
at $(\beta ,u)=(4.8, 1.56)$ and $(5.0, 1.5)$ in simulations on an
$8^3\times 16$ lattice, while it is absent at $(4.8, 1.5)$, $(5.0,
1.48)$, $(5.2, 1.44)$ and $(5.3, 1.33)$, see Figure~\ref{figPa}.
However these results are obtained with limited statistics and may
be therefore influenced by insufficient thermalisation. Also at
$\beta=5.0, u=1.5$ the number of CG iterations along the plateau which
originated from the hot start is around $500$ whereas it drops to
$350$ when started from a unit configuration. This may suggest
that there may be a difference in the pion mass for these two
phases. We also note that on the finite temperature lattice,
$8^3\times 4$, the meta stability signal has disappeared at
$\beta=5.0, u=1.5$. From this one may conclude that if present the
first order phase transition becomes softer and is shifted in the
direction of smaller current quark mass respectively larger $u$
when the temperature is raised. The presence of a meta stability
region sets a lower bound on the pion mass below which one cannot
relate the results at given values of $\beta$ and $u$ to the
correct universality class of the physical theory. Otherwise one
could go to quite small pion masses in the region below
$\beta=5.0$.

\begin{figure}[h]
\centering
\includegraphics[angle=-90,scale=0.55]{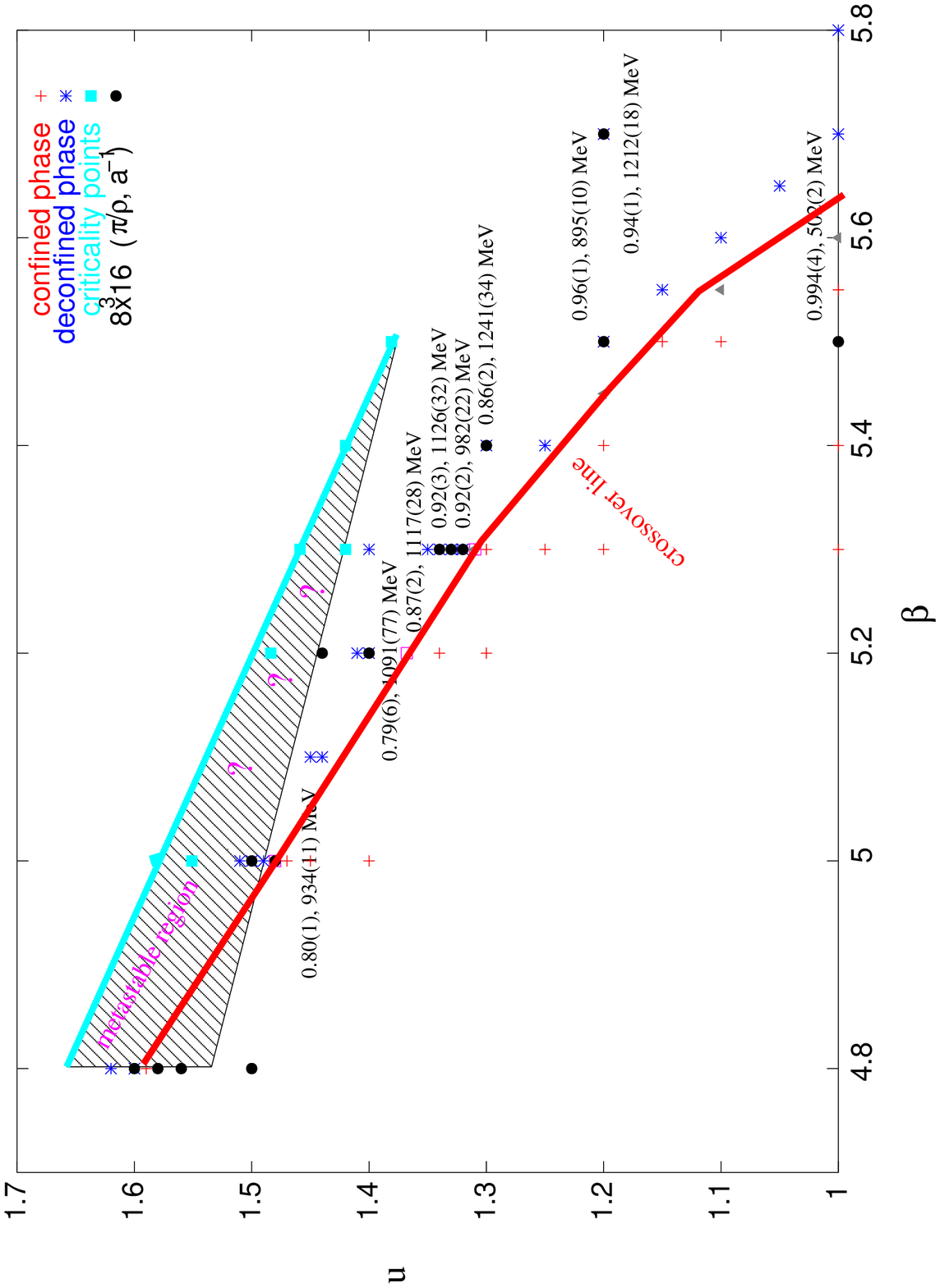}
\caption{  Conjectured phase diagram determined by inspecting the
Polyakov loop and the inverse number of iterations required by a
CG solver in the course of the molecular dynamics is shown. In
particular the position of the critical line, the thermal
crossover line and the region where meta stability may occur is
shown. The black dots indicate simulations on $8^3\times 16$
lattices together with the ratio of the pseudoscalar and the
vector meson mass and an estimated inverse lattice spacing. The
error bars are purely statistical.} \label{figPD}
\end{figure}

In Figure~\ref{figPD} we conjecture the structure of the lattice
phase diagram $\{\beta,u\}$ for an $8^3\times 4$ lattice based on
the current data. The red line denotes the crossover from the
confined into the deconfined phases, blue line shows approximately
where the critical line is and the shaded region indicates meta
stability.
\section{Conclusions}
We simulated two flavours of dynamical fermions with a truncated
perfect action implementing the exact fermionic force in the
Hybrid Monte Carlo. For the gauge action we used the conventional
plaquette action. So far we do not observe the Aoki phase at the
investigated parameters. Instead, we observe some indications for
the existence of a meta stable region in our simulations on an
$8^3\times 16$ lattice. This artifact may disappear at the same
parameter values on an $8^3\times 4$ lattice, which would signal
that this region may be shifted in the direction of smaller pion
masses once the temperature is increased. The presence of the meta
stable region may prevent us from studying the critical
temperature at small pion masses. Hopefully this problem will
soften as one applies a different gauge action, see
e.g.~\cite{AliKhan:2000iz,Farchioni:2004fs}, or uses some variants
of the smearing in the fermionic action, see
e.g.~\cite{Morningstar:2003gk}. Also one may consider to simulate
an approximate overlap operator with the kernel of the Hypercube
operator.
 The smallest confirmed ratio of the pseudoscalar mass to the vector meson
mass at the crossover line at $N_t=4$ is around $0.8$. Here the
estimate for the critical temperature is about $234 \pm 3$ MeV
which roughly agrees with the world data at this pion mass. Given
the fact that this ratio is realised at $N_t=4$ the Hypercube
operator is still cheaper than the Wilson operator where such
ratio was reported only at $N_t=8$ using the plaquette gauge
action, see e.g.~\cite{Bornyakov:2005dt}.

\acknowledgments{ S.S. is indebted to Wolfgang Bietenholz for
numerous discussions and comments on the manuscript and to Jan
Volkholz for providing me with a code for the Sexton-Weingarten
integrator. S.S. also acknowledges discussions with S. D\"urr, S.
Ejiri and F. Karsch during the lattice conference. Most
computations were made on the supercomputer centre "Norddeutscher
Verbund f\"ur Hoch und H\"ochstleistungsrechnen"~(HLRN).}

\end{document}